\documentclass[useAMS,usenatbib,usegraphicx]{mn2e}			
\usepackage{times}
\usepackage{natbib} 
\usepackage[dvips]{graphicx,color}
\usepackage{amsmath}
\usepackage{amssymb}

\title[Disentangling the stellar populations in
NGC~4550]{Disentangling the stellar populations in the
  counter-rotating disc galaxy NGC~4550}
\author[E.J. Johnston et al.]{Evelyn~J.~Johnston,$^1$\thanks{Email: ppxej@nottingham.ac.uk} Michael~R.~Merrifield,$^1$ Alfonso~Arag\'on-Salamanca$^1$ and \newauthor Michele~Cappellari$^{2}$\\
  $^1$School of Physics and Astronomy, University of Nottingham, University Park, Nottingham, NG7 2RD, UK\\
  $^2$Sub-department of Astrophysics, Department of Physics,
  University of Oxford, Denys Wilkinson Building, Keble Road, Oxford,
  OX1 3RH, UK}

\begin{document}

\maketitle

\begin{abstract}
  In order to try and understand its origins, we present high-quality
  long-slit spectral observations of the counter-rotating stellar
  discs in the strange S0 galaxy NGC~4550.  We kinematically decompose
  the spectra into two counter-rotating stellar components (plus a
  gaseous component), in order to study both their kinematics and
  their populations.  The derived kinematics largely confirm what was
  known previously about the stellar discs, but trace them to larger
  radii with smaller errors; the fitted gaseous component allows us to
  trace the hydrogen emission lines for the first time, which are
  found to follow the same rather strange kinematics previously seen
  in the [OIII] line.  Analysis of the populations of the two separate
  stellar components shows that the secondary disc has a significantly
  younger mean age than the primary disc, consistent with later star
  formation from the associated gaseous material.  In addition, the
  secondary disc is somewhat brighter, also consistent with such
  additional star formation.  However, these measurements cannot be
  self-consistently modelled by a scenario in which extra stars have
  been added to initially-identical counter-rotating stellar discs,
  which rules out Evans \& Collett's (1994) elegant
  ``separatrix-crossing'' model for the formation of such massive
  counter-rotating discs from a single galaxy, leaving some form of
  unusual gas accretion history as the most likely formation
  mechanism.
\end{abstract}

\begin{keywords}
  galaxies: elliptical and lenticular -- galaxies: evolution --
  galaxies: formation -- galaxies: individual (NGC~4550) -- galaxies:
  kinematics and dynamics -- galaxies: stellar content
\end{keywords}

\section{Introduction}\label{sec:introduction}
The presence of counter-rotating populations of stars in S0 galaxies
is a well-known phenomenon, although reasonably uncommon: while almost
a quarter of S0 galaxies contain gaseous components that
counter-rotate relative to their stars, less than 10\% were found to
contain distinct counter-rotating stellar disc components 
\citep{Bertola_1992,Kuijken_1996}. However, such systems do exist, and so any
theory of S0 formation must provide a channel for their creation.

Of these counter-rotating systems, the one that presents the greatest
challenge to models of galaxy formation is
NGC~4550. \citet{Rubin_1992} obtained long-slit spectra along the
major axis of this normal-looking S0 galaxy, and found that the
absorption-lines split neatly into two, indicating two extended
counter-rotating discs. Subsequent analysis has confirmed that these 
discs are very similar, with comparable sizes, masses, kinematics 
and line strength \citep{Rix_1992}. Integral-field observations, 
in combination with dynamical models, confirmed the picture, but 
revealed some breaking of the symmetry, with one disc being thicker 
than the other \citep{Cappellari_2007} and also containing an 
emission-lines component \citep{Sarzi_2006}.

The reason that it is difficult to come up with a scenario for
constructing such a system is that the obvious solution of merging two
normal discs with opposite angular momenta does not generally work.
In particular, it has long been known that most mergers between two
roughly equally-massive discs are likely to be very destructive,
heating the system enormously and not resulting in the required
disc-like final morphology \citep{Toomre_1977}.  However, this problem
may not be insurmountable: \citet{Puerari_2001} show that a major
merger between disc galaxies of comparable mass could produce the
kinematics seen in NGC~4550, as long as the initial conditions are
just right, with the precursor systems co-planar on a carefully-chosen
parabolic orbit.  A similar result was found by \citep{Crocker_2009}, 
who also tried to reproduce the gas kinematics of the galaxy.
Clearly, such an arrangement is rather contrived, but not impossible 
if systems like NGC~4550 are truly rare. The ATLAS$^{\rm 3D}$ survey 
\citep{Cappellari_2011} obtained integral field stellar kinematics 
for a volume-limited sample of 260 early-type galaxies. Out of these 
they found 11 cases (see fig.~C5 of \citealt{Krajnovic_2011}) showing evidence 
for major counter-rotating stellar discs. However in only about half of 
these (about 2\% of the sample) the two counter-rotating discs seem 
to have comparable mass like NGC~4550.

An alternative scenario that avoids the destructive force of a major
merger is the possibility that counter-rotating gas could be accreted
rather slowly by a normal disc galaxy, and subsequently form stars in
a new counter-rotating disc.  This possibility was explored through
simulations by \citet{Thakar_1996,Thakar_1998}, who found that the
counter-rotating stellar disc formed in this way tended to be rather
small.  They did find that a series of mergers with gas-rich dwarf
galaxies could produce a counter-rotating disc of comparable mass and
size to the original, but once again the initial conditions needed to
be very carefully tuned to produce such matched discs.  

The similarity of the two discs in NGC~4550 led \citet{Evans_1994} to
suggest a third possibility that could much more naturally produce
identical counter-rotating discs.  In this 
``separatrix-crossing'' scenario, a single initially-triaxial
elliptical galaxy evolves slowly with time into an axisymmetric
morphology.  At that point, the family of box orbits that existed in
the triaxial system would disappear, and stars would switch onto tube
orbits instead.  Since the initial box orbit had no preferred sense of
rotation, stars would end up randomly on tube orbits rotating in
either sense around the centre of the now-axisymmetric system, thus
automatically generating a pair of identical counter-rotating stellar
populations.

To-date, it has not been clear which, if any, of these scenarios might
be responsible for the formation of NGC~4550.  However, they do
predict some distinct difference in the resulting counter-rotating
discs, particularly in the properties of their stellar populations,
which we might be able to use to distinguish between them.  While the
separatrix-crossing scenario will produce truly identical discs, a
counter-rotating stellar disc formed by gas accretion must have a
younger population than the pre-existing disc, while the
counter-rotating discs in a system formed by a merger will reflect the
stellar populations of the progenitor galaxies.

In this paper, we set out to analyse simultaneously the kinematics and
stellar populations of NGC~4550, to try to distinguish between these
possibilities.  The remainder of the paper is laid out as follows:
Section~\ref{sec:Observations and Data Reduction} describes the new
spectral data we have used; in Section~\ref{sec:Kinematics}, we
present the fitting technique developed to separate out the stellar
component, and present the resulting kinematic measurements;
Section~\ref{sec:Stellar Populations} analyses the individual
component spectra to quantify their populations; and
Section~\ref{sec:Discussion and Conclusions} discusses the
implications of these measurements for the various theories as to how
this galaxy might have formed.

\section{Observations and Data Reduction}\label{sec:Observations and Data Reduction}
The requisite spectral observations along the major axis of NGC~4550
were carried out using the GMOS instruments in long-slit mode on
Gemini-North and Gemini-South on 2009 June 20 and 2010 February 13
respectively. 1800 seconds of exposure was obtained with each
telescope.  Spatially, the chips were binned by 4, to give a final
scale of 0.29 arcseconds pixel$^{-1}$.  The measured seeing of
$\sim1.77$~arcsec with Gemini-South and $\sim{0.9}$~arcsec with
Gemini-North was well below the spatial scales of interest in this
extended system -- adopting a distance of $15.5\,{\rm Mpc}$
to this galaxy \citep{Mei_2007}, $1\,{\rm arcsec} \equiv 75\,{\rm
  pc}$.  We used a 0.5 arcsecond slit and the B1200 grating, with a
central wavelength of $\sim 4730$\AA, slightly offset between the two
sets of exposures to fill in the gaps between the chips.  These
observations thus provided an hour of integration over the spectral
range $4100-5450$\AA\ with a dispersion of 0.235\AA\ pixel$^{-1}$. The
spectral resolution was measured from the FWHM of the arc lines to be
$\sim{1.13}$\AA, which corresponds to a velocity resolution of
$72\,{\rm km}\,{\rm s}^{-1}$ FWHM or a velocity dispersion of
$30\,{\rm km}\,{\rm s}^{-1}$.

As part of a larger programme of observations, a series of
spectrophotometric and template stars were also observed with the same
instrumental set up, of which the details are given in
Table~\ref{star_info}. The template stars cover a range of spectral
types in order to match the composite spectral type of the galaxy in
our combined stellar kinematic and population analysis.

\renewcommand{\tabcolsep}{0.67cm}
\begin{table}
\begin{center}
\caption{Spectrophotometric (S) and template (T) stars.\label{star_info}}
\begin{tabular}{ l l l}
\hline \hline
 Name & T/S  & Spectral Class \\
\hline
HD054719 & T & K2 III \\
HD070272 & T & K5 III \\
HD072324 & T & G9 III \\
HD073593 & T & G8 IV \\
HD120136 & T & F6 IV \\
HD144872 & T & K3 V \\
HD145148 & T & K0 IV \\
HD161817 & T & A2 VI \\
Feige66 & S & - \\
Hiltner600 & S & - \\
LTT1788 & S & - \\
\hline
\end{tabular}
\end{center}
\end{table}

The science spectra, along with dome flats taken following each set of
observations and CuAr arc spectra from the same instrumental set-up,
were reduced using the GMOS spectral reduction packages in
\textsc{iraf}.\footnote{ \textsc{iraf} is distributed by the National
  Optical Astronomy Observatories, which are operated by the
  Association of Universities for Research in Astronomy, Inc., under
  cooperative agreement with the National Science Foundation} All the
science and calibration frames were reduced by applying bias
subtraction, flat fielding, cosmic ray removal and an initial
wavelength calibration, and the three sections of each spectrum from
each CCD were joined together. The arc spectra were then used to
correct for the geometric distortions caused by the instrument optics
and to refine the wavelength solution over the whole spectrum; the
residuals of the resulting wavelength fits were $\sim 0.2-0.3$ \AA.

The wavelength-calibrated spectra were then sky subtracted, corrected
for atmospheric extinction and flux calibrated using the
spectrophotometric standard star spectra.  Finally, the spectra from
each telescope were combined, using the measured positions of
prominent sky lines to ensure the best possible registration of
absolute wavelength calibration.  The resulting two-dimensional
spectrum for the galaxy, representing an hour of integration on an
8-metre telescope, provides the very high signal-to-noise ratio
required for this combined kinematic and population analysis: the
central 2 arcseconds of the galaxy yield a spectrum with a
signal-to-noise ratio in excess of 100, and judicious binning allows
us to keep the signal-to-noise ratio in excess of 20 throughout the
galaxy.

\begin{figure*}
  \includegraphics[width=1\linewidth]{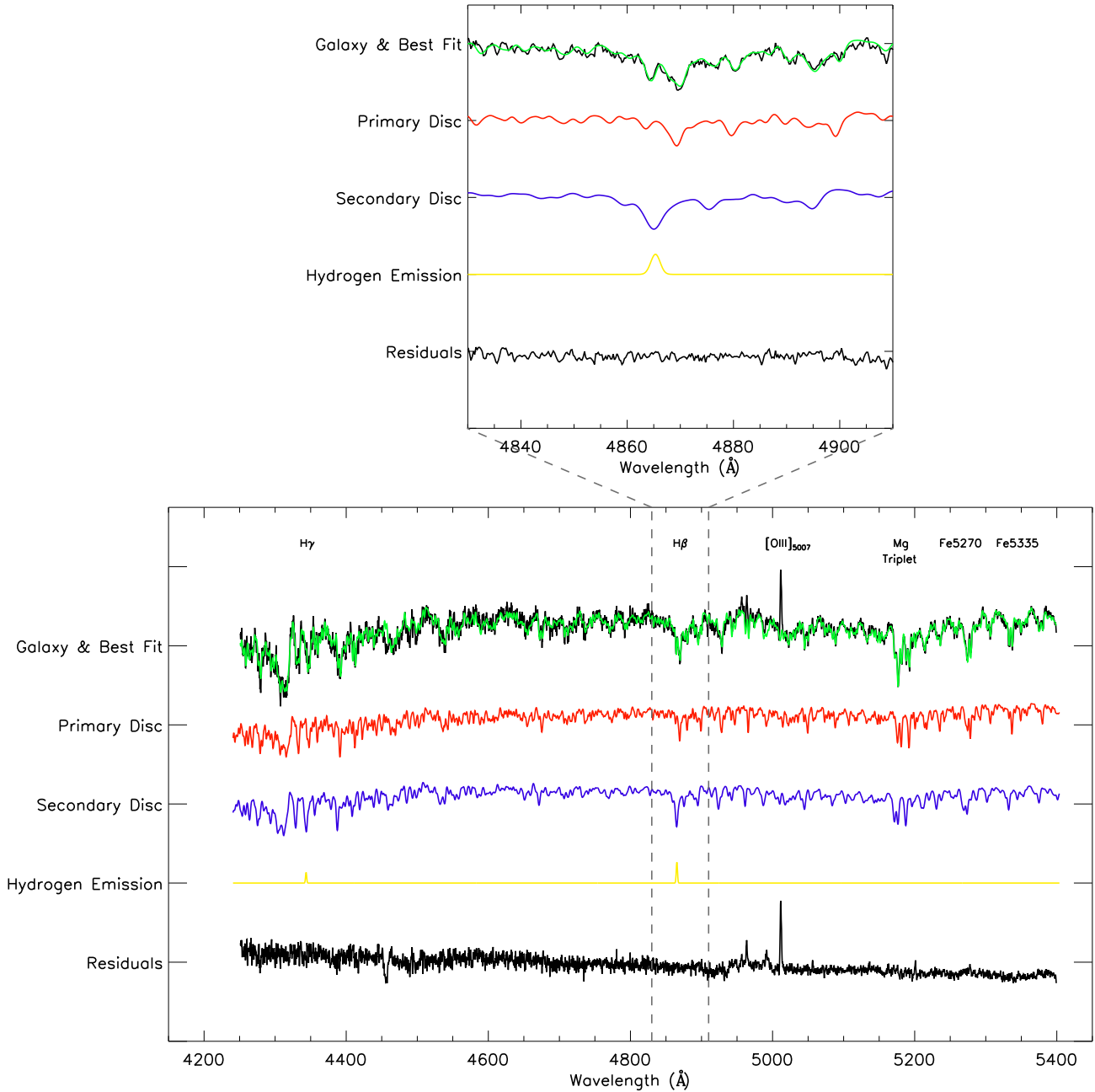}
  \caption{The spectrum of one side of the outer disc of NGC~4550,
    showing both the full spectral range and a zoom in on the H$\beta$
    line.  The data are shown in black, and the green line shows the best-fit
    model.  The individual components that comprise this model and the
    residuals of the fit are also shown. Note- the polynomials are not 
    included in this fit. \label{components}}
\end{figure*}

\begin{figure*}									
  \includegraphics[width=1\linewidth]{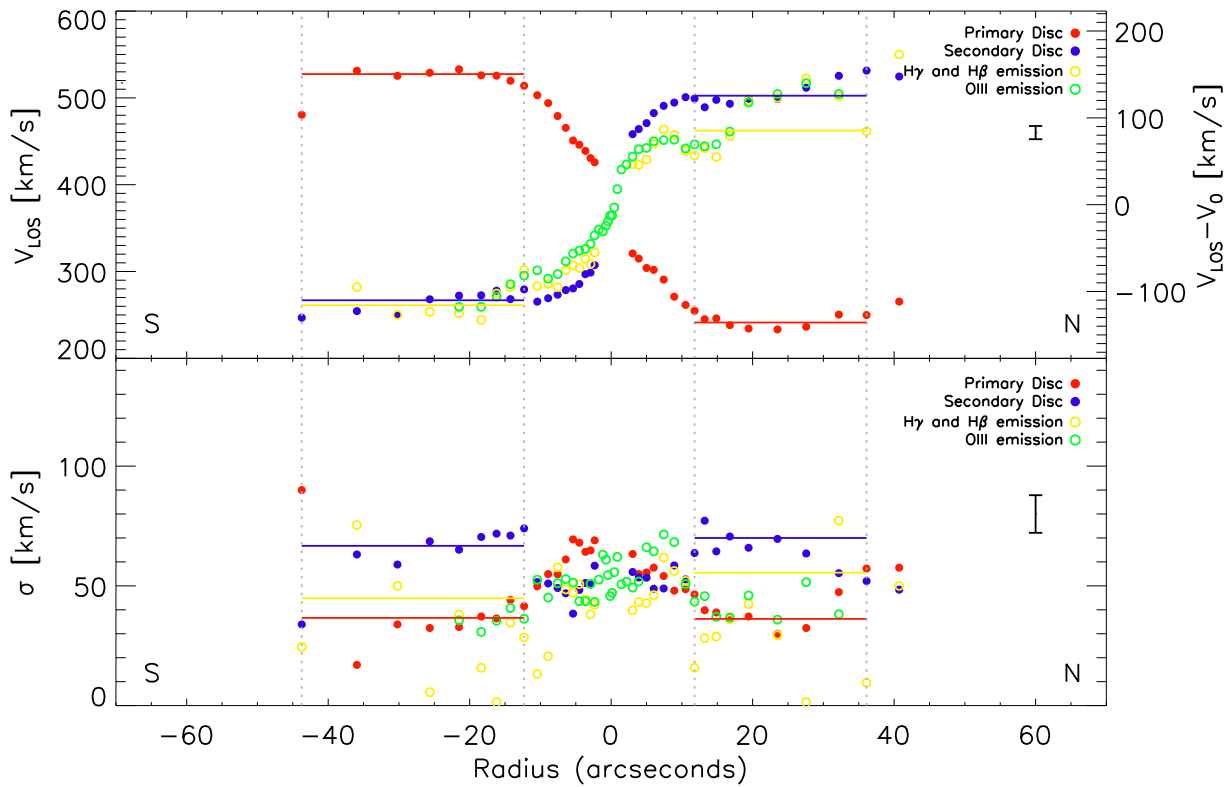}
  \caption{Measurements of the radial velocity and velocity dispersion for each 
	kinematic component fitted to the spectra of NGC~4550 as a
        function of radial position along the major axis (with north
        and south directions annotated).  Typical
        characteristic error bars are shown on the right of each plot.
        The horizontal lines show the results of fitting to spectra
        co-added over the entire flat part of the rotation curve
        (marked by vertical dotted lines).  The
        kinematics of the [OIII]$_{\lambda5007}$ line, derived by
        direct fitting to the spectral line, are also shown.  
        \label{kinematics_ELODIE}}
\end{figure*}

\section{Kinematic Decomposition}\label{sec:Kinematics}

In order to study the stellar populations of the two component
counter-rotating discs, we must first separate their
spectra. Fortunately, outside the central few arcseconds, the
line-of-sight velocities of the two components are different enough to
split the corresponding absorption lines quite cleanly -- indeed, it
was this splitting that enabled \citet{Rubin_1992} to identify the
counter-rotating discs in the first place -- so we can fit a
two-component spectral model, where each component has a different
mean velocity and velocity dispersion, reasonably unambiguously.  At
the same time, we also have to allow for the unknown stellar
population properties of the two components.  In order to fit all
these factors simultaneously, we modified the Penalized Pixel
Fitting code (\textsc{ppxf}) of \citet{Cappellari_2004} in a similar
way to \citet{Coccato_2011}. This code combines the template stars
listed in Table~\ref{star_info} to produce two model spectra
representing the two stellar population components, which when added
together would best fit the galaxy spectrum. To achieve the best fit,
the component spectra were multiplied by low-order Legendre
polynomials to model out any mismatch in the flux calibration of the
continuum, and convolved with line-of-sight velocity distributions of
different shapes, mean velocities and dispersions, to best represent
the kinematics of each component.

One further complication is that it is clear from the raw spectra that
there is a third, gaseous, component rotating in one direction, as is
evident from the strong [OIII]$_{\lambda5007}$ emission line.  We can
deal with this contaminant by simply masking it from the spectral
range used in the fitting.  However, there is, presumably, also
emission from hydrogen gas, which usually accompanies the [OIII] line
in galaxies.  Such emission could prove disastrous for the analysis of
this paper, as it would partially fill in the hydrogen absorption
lines of one of the stellar components, completely altering the
derived line strengths of these important lines, and thus the inferred
properties of the stellar population.  We therefore fit a third
component consisting simply of two Gaussians at the wavelengths of the
$H\beta$ and $H\gamma$ features, with a FWHM equal to the spectral
resolution of the galaxy spectrum and the ratio of their intensities
given by the Balmer decrement from \citet{Raynolds_1997}.  By
convolving this component with its own velocity distribution in the
fit process, we also obtain the kinematics of the gas component.
Fortunately,due to the long wavelength range over which the 
spectra are fit, and because the ratio of hydrogen emission line strengths in
the gaseous component is different from the ratio of the corresponding
absorption lines in the underlying stellar component, there is not a
degeneracy in the resulting fit, so both components can be
independently extracted. In principle, differential reddening 
could introduce a degeneracy problem, but because this galaxy is an S0
with little evidence for large amounts of dust, 
the effect is not thought to be significant.

A typical result of this fitting process is illustrated in
Figure~\ref{components}.  As well as confirming the generally very
good job that this fitting process does in reproducing the full
spectrum with the three components, this figure also underlines the
importance of including the gas component: from the raw spectrum, one
might conclude that the H$\beta$ absorption line is somewhat stronger
in the redshifted (primary disc) component, but the full fitting
process reveals that this conclusion is driven by the filling in of
the absorption feature in the blueshifted component by the emission
line, and actually it is this secondary-disc component that has the
stronger absorption line.  

\begin{figure*}
  \includegraphics[trim=0cm 10cm 0cm 0cm, clip=true,width=0.9\linewidth]{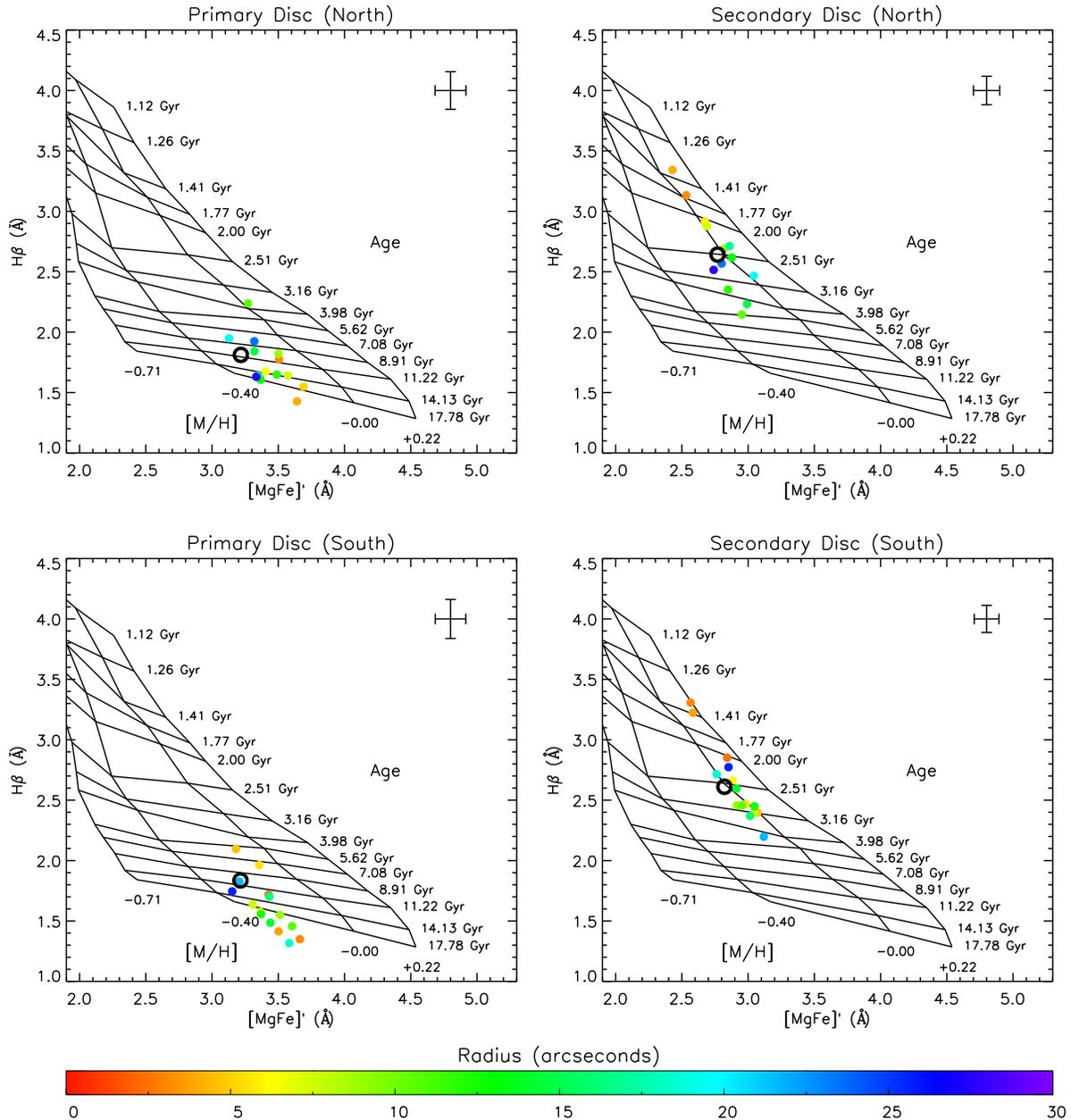}
  \caption{The line indices derived from the model components compared
	to the predictions of SSP models, measured out to a 
	radius of $\sim$~30~arcsec, or $\sim$~2.25~kpc, where the S/N  
	drops below 30~per~Angstrom.  The primary and 
	secondary discs are in the left and right columns, while the
        north and south side of the galaxy are in the upper and lower
        rows.   The radius of each measurement is color-coded; the
        open black points show the average value for the outer part of
        each disc.  For clarity, points are plotted without errors; a
        typical error bar is shown in the top right of each plot.  
        \label{SSP_models}}
\end{figure*}

This fitting process was repeated using the spectral data from all
along the major axis, co-added spatially to maintain a signal-to-noise
ratio of at least 20 per pixel.  Only the central $\sim$~5~arcsec
could not be decomposed in this way, due in combination to the overlapping 
kinematics of the two discs and the increasing contribution to the light 
from the bulge. The mean velocities and velocity
dispersions derived for the three components are shown in
Figure~\ref{kinematics_ELODIE}.  Errors on each point were estimated
by Monte Carlo simulations of model galaxies constructed using the
same components as in the fit; for clarity, we do not plot all points
with error bars, but show the mean resulting error on the right of the
plot.  We also tested the sensitivity of the kinematic results to the
spectral templates adopted by repeating the analysis using a sub-sample
of the \textsc{elodie} spectral templates \citep{Prugniel_2001} that 
covered the full range of spectral classes, but found
no significant systematic differences.  Co-adding the data from the
flat part of the rotation curve, we find a primary disc with a
rotation velocity of $143\pm7$~km~s$^{-1}$ and a velocity dispersion of
$36\pm7$~km~s$^{-1}$, and a secondary disc with a lower rotation
velocity of $-118\pm8$~km~s$^{-1}$ and a higher velocity dispersion of
$68\pm10$~km~s$^{-1}$, consistent with the previous findings of
\citet{Rix_1992}.

There is clearly something a little strange about the kinematics
derived for the emission-line gas disc.  Because the gas shows a lower
velocity dispersion than the secondary stellar disc that co-rotates
with it, one would expect it to display a smaller amount of asymmetric
drift, and hence rotate more quickly, whereas it actually rotates 
slower than the accompanying stars.  This strange property does not
seem to be the result of any failing in extracting the gas kinematics
correctly: we can obtain some confidence that the fitting process is
picking up the correct properties for the hydrogen emission lines by
comparing the results obtained to those measured directly from the
excluded [OIII]$_{\lambda5007}$ line, also shown in
Figure~\ref{kinematics_ELODIE}, which are clearly very similar.  The
most likely explanation is therefore that the gas does not form a
simple equilibrium axisymmetric disc, and hence would not obey the
usual asymmetric drift equation. Some indication of an asymmetry in the 
[OIII] gas velocity with respect to the projected major axis, is visible 
in \citet{Sarzi_2006}.

\section{Stellar Populations}\label{sec:Stellar Populations}
Having kinematically decomposed the spectra into the two stellar disc
components at each radius, we can now study the stellar populations of
each individual component, as derived from the strengths of its
absorption lines.  In order to render the line strengths at different
radii comparable, all component spectra were broadened with Gaussians
to match their dispersions to the largest values found at the centre
of the galaxy.  The H$\beta$, Mgb, Fe5270 and Fe5335 indices were then
measured using the \mbox{\textsc{indexf}}
package\footnote{http://www.ucm.es/info/Astrof/software/indexf/indexf.html}.
This software uses the Lick/IDS index definitions to calculate a
pseudo-continuum over each absorption feature based on the level of
the spectrum in bands on either side, and measures the strength of the
feature relative to the pseudo-continuum \citep{Worthey_1994,
  Worthey_1997}. The combined metallicity index, [MgFe]$'$
\citep{Gonzalez_1993,Thomas_2003}, was then determined from these
values.  The uncertainty in each measurement was estimated from the
propagation of random errors and the effect of uncertainties in the
radial velocity.

Figure~\ref{SSP_models} shows the resulting values of the H$\beta$
index, as an indicator of stellar population age, plotted against the
[MgFe]$'$ index, as an indicator of metallicity, for each component as
a function of radius.  We also show the average results obtained by
combining all the outer disc data from the flat part of the rotation
curve, as delineated in Figure~\ref{kinematics_ELODIE}.  Since we have
observations of both sides of the galaxy, in which the
counter-rotating components will be Doppler shifted in opposite
directions, we obtain two independent measurements of these quantities
with potentially different systematic biases as different spectral
features in the two components will end up superimposed in the
composite spectra from each side.  The good agreement between the two
rows of plots shown in Fig.~\ref{SSP_models} again provides some
confidence in the results.  What is most striking about these plots is
the systematic difference between the H$\beta$ indices between the two
components.  In addition, there seems to be a fairly strong gradient
in the H$\beta$ index of the secondary component, with the largest
values at the smallest radii.

The H$\beta$ index is largely a measure of the age of the stellar
population, and we can quantify what the differences mean by
calculating simple stellar population (SSP) models for comparison with
the data.  Here, we have used the web
interface\footnote{http://miles.iac.es/} to the models of
\citet{Vazdekis_2010}, which use the \textsc{miles} stellar library
\citep{Sanchez_2006}, and degrade the library spectra to match the
resolution of the data by convolving them with a Gaussian of the
appropriate dispersion.  In this way, we can obtain estimates of the
relative ages and metallicities without the potential loss of
information that would normally result from degrading the galaxy
spectra and models right down to the resolution of the Lick indices.
Figure~\ref{SSP_models} shows the line indices for the resulting grid
of models of different ages and metallicities.  Although the usual
caveats apply to the absolute values of physical parameters derived
from such analyses, it is clear that the two discs have systematically
different age properties.  In particular, the primary disc is old,
with an age inferred here of $\sim 11\,{\rm Gyr}$, while the secondary
disc, which rotates in the same direction as the gas, is much younger
at $\sim 2.5\,{\rm Gyr}$, with its innermost parts even more youthful.
Interestingly the young component also co-rotates with the molecular 
gas \citep{Crocker_2009}, which is generally associated to recent 
star formation episodes. 

\section{Discussion and Conclusions}\label{sec:Discussion and Conclusions}

By carefully disentangling the spectral components of NGC~4550, we
have been able to learn a great deal about their individual
properties, determining quantities that have significant implications
for how this peculiar system might have formed.  In particular, the
strong differences in the stellar populations of the two discs seem
to rule out the separatrix-crossing model in which they were formed
from a single parent stellar population.  It is interesting that one
of the things that motivated \citet{Evans_1994} to consider this model
in the first place was the apparent similarity of line strengths
between components; as we have now seen, this similarity in apparent
H$\beta$ line strengths arises from an unfortunate cancellation
between the stronger absorption lines of a younger population
superimposed on the emission lines from the gas that rotates in the
same direction.

\begin{figure}									
  \includegraphics[width=1\linewidth]{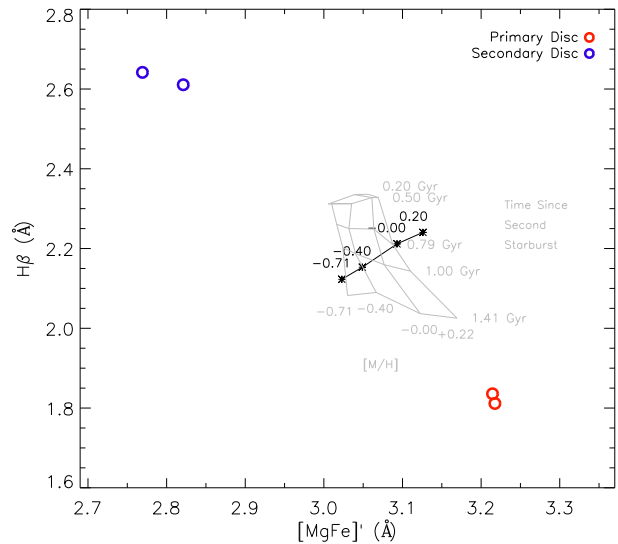}
  \caption{Model line indices derived for different star formation
    histories following separatrix crossing.  The black line shows the
    effect on the line indices of the primary disc if the initial
    population had continuous steady star formation of differing
    metallicities until the present day, while the grey grid shows the
    effect of a single burst of star formation at the times and
    metallicities indicated. The open circles are the results
    for the two sides of the primary and secondary discs from
    Fig.~\ref{SSP_models}.
    \label{SSP_model_2starburst}}
\end{figure}

However, the very presence of this gas suggests that all may not be
lost for the separatrix-crossing model.  Perhaps this scenario did
indeed occur, creating two initially-identical stellar discs.
Subsequent accretion then created a gas disc rotating in the direction
of one of the stellar discs, and this gas then formed further
generations of stars, creating a composite population in the disc
co-rotating with the gas whose mean age would appear younger, as
observed.  Indeed, one could always invoke sufficient recent star
formation to shift the inferred age from $\sim 11\,{\rm Gyr}$ to $\sim
2.5\,{\rm Gyr}$.  Since the line-index ages are effectively
luminosity-weighted, a relatively modest amount of star formation,
creating a bright young population, might explain the observed age
differential.

Fortunately, we have one further constraint from this analysis that we
can use to assess the viability of this modified scenario.
Specifically, the decomposition of the spectra into the two stellar
discs also tells us how much total light should be attributed to each
component.  Clearly, if one of the two initially-identical discs has
had significant new stars forming in it, this component will have a
greater luminosity.  Encouragingly, this is what we find: the
decomposition of the whole outer disc spectra (the open symbols in
Figure~\ref{SSP_models}) reveals that the secondary component, which
co-rotates with the gas, has a continuum level at 4400\AA\ (the centre
of the B-band) that is 20\% higher than the gas-free primary
component.  

So now we have an extra constraint which means that the amount of late
star formation we can add is fixed by this additional 20\% of B-band
luminosity.  We do not know the exact star formation history of any
such late addition, but we can try out different possibilities.
Figure~\ref{SSP_model_2starburst} shows the results of such attempts
again using the \citet{Vazdekis_2010} models.  The points at the
bottom right show the line indices measured for the primary disc,
which form the presumed starting point of the stellar population of
both discs at the moment of separatrix crossing.  The points at the
top left show the higher indices that we are trying to reach by adding
subsequent star formation.  The line of crosses show what happens if
we invoke continuous steady star formation of different metallicities
ever since the old stellar population formed: the level of this star
formation is then uniquely fixed by the requirement that its addition
results in an enhancement of the disc's total B-band luminosity by
20\%.  Similarly, the grid of points shows the change in line indices
caused by the addition of a single burst of star formation of varying
ages and metallicities, again with the amplitude of the burst tuned to
match the enhanced total luminosity of the disc.  Since these two
extreme possibilities of star formation history move the disc to the
same region of the plot, it is not surprising that other more
complicated possibilities also all end up in the same area.  Clearly,
although this additional later star formation moves the line indices
in the right direction, it is nowhere near sufficient to reproduce the
observed values for the younger disc.

With regret, we are therefore forced to abandon the elegance of the
separatrix-crossing model entirely, and conclude that NGC~4550 formed
through one of the other scenarios.  The ages inferred in
Section~\ref{sec:Stellar Populations} then tell us something about the
process.  If formed through a carefully-controlled merger of
fully-formed galaxies, these ages just reflect the ages of the
progenitors.  In the gas-rich accretion scenario, which now seems more
natural, the $\sim 2.5\,{\rm Gyr}$ age of the secondary disc tells us
how long ago this gas was accreted, with the residue of this accreted
gaseous material still rotating along with this component, albeit in a
somewhat non-circular manner.  The higher velocity dispersion of this
younger disc then presumably reflects the more turbulent nature of
such secondary gas accretion when compared to the more conventional
formation of the older primary stellar disc.  We even begin to obtain
some insight into the spatial distribution of this star formation,
with the age gradient in the secondary component implying that the
star formation has become ever more centrally concentrated as the gas
has been depleted.  If this scenario is correct, the only unexplained
phenomenon is why the two counter-rotating disc components have such
similar spatial extents, which at this point we must simply attribute
to coincidence.

\section*{Acknowledgements}
    We would like to thank Bruno Rodriguez del Pino for his help in 
    debugging the modified code and for several useful discussions. 
    We would also like to thank the anonymous referee for their 
    useful comments that helped improve this paper.
    This work was based on observations obtained at the Gemini Observatory, 
    which is operated by the Association of Universities for
    Research in Astronomy, Inc., under a cooperative agreement with the
    NSF on behalf of the Gemini partnership: the National Science
    Foundation (United States), the Science and Technology Facilities
    Council (United Kingdom), the National Research Council (Canada),
    CONICYT (Chile), the Australian Research Council (Australia),
    Minist\'{e}rio da Ci\^{e}ncia, Tecnologia e Inova\c{c}\~{a}o (Brazil)
    and Ministerio de Ciencia, Tecnolog\'{i}a e Innovaci\'{o}n Productiva
    (Argentina). The programme IDs were GN-2009A-Q-102 and GS-2010A-Q-23.
    EJ acknowledges support from STFC, and MC acknowledges support from 
    a Royal Society University Research Fellowship

\bibliographystyle{mn2e}

\bibliography{4550_refs}

\end{document}